\documentclass{aa}
\usepackage{natbib,amsmath,amssymb,graphicx}
\newcommand{\rsun}{R_{\odot}}

\begin{document}

\title{Modeling Solar Cycles 15 to 21 Using a Flux Transport Dynamo}

\author{J. Jiang\inst{1}
 \and R.H. Cameron\inst{2}
 \and D. Schmitt\inst{2}
 \and E. I\c{s}\i k\inst{3}}

\institute{Key Laboratory of Solar Activity, National
Astronomical Observatories, Chinese Academy of Sciences, Beijing
100012, China \email{jiejiang@nao.cas.cn}
\and Max-Planck-Institut f\"ur Sonnensystemforschung,
  37191 Katlenburg-Lindau, Germany
\and Department of Physics, Faculty of Science \&
Letters, \.Istanbul K\"ult\"ur University, Atak\"oy Campus,
   Bak\i rk\"oy 34156, \.Istanbul, Turkey}
\date{Received ; accepted}

\abstract {The Sun's polar fields and open flux around the time of
activity minima have been considered to be strongly correlated with
the strength of the subsequent maximum of solar activity.} {We aim
to investigate the behavior of a Babcock-Leighton dynamo with a
source poloidal term that is based on the observed sunspot areas and
tilts. In particular, we investigate whether the toroidal fields at
the base of convection zone from the model are correlated with the
observed solar cycle activity maxima.} {We used a flux transport
dynamo model that includes convective pumping and a poloidal source
term based on the historical record of sunspot group areas,
locations, and tilt angles to simulate solar cycles 15 to 21.} {We
find that the polar fields near minima and the toroidal flux at the
base of the convection zone are both highly correlated with the
subsequent maxima of solar activity levels ($r=0.85$ and $r=0.93$,
respectively).} {The Babcock-Leighton dynamo is consistent with the
observationally inferred correlations.}

\authorrunning{Jiang et al.}
\titlerunning{Modeling Solar Cycles 15 to 21 Using a Flux Transport Dynamo}
\maketitle

\section{Introduction}
The solar magnetic cycle is believed to be the result of a
magnetohydrodynamic dynamo. One family of these dynamo models is the
Babcock-Leighton (BL) model, which was first proposed by
\cite{Babcock61} and further elaborated on by \cite{Leighton64}. The
essence of the BL mechanism is that the generation of the poloidal
field comes from the decay of the tilt sunspot groups on the solar
surface. BL models have been successful in reproducing some
characteristics of the solar cycle \citep[e.g.,][]{
choudhuri95,Durney97,Choudhuri99,Dikpati99,Nandy02,Chatterjee04,Rempel06,Guerrero07,Yeates08},
including the solar cycle irregularities \citep[e.g.,][]{
Charbonneau00,Durney00,Charbonneau01,Karak10,Karak11}. Recent
analyses of the long-term sunspot observations strengthen the idea
that the dynamo is consistent with the BL mechanism
(\citeauthor{Dasi10}, \citeyear{Dasi10}; \citeauthor{Cameron10},
\citeyear{Cameron10} [hereafter CJSS10]; \citeauthor{Kitchatinov11},
\citeyear{Kitchatinov11}).

The important ingredients in the BL dynamo model are ({\it{i}}) the
generation of poloidal flux due to the emergence of tilted sunspot
groups and the subsequent evolution of the surface magnetic flux,
({\it{ii}}) the transport of poloidal flux to the tachocline,
({\it{iii}}) the generation of toroidal magnetic flux due to the
winding up of the poloidal field (the $\Omega$-effect), and
({\it{iv}}) the subsequent formation and rise of flux loops through
the convection zone until they emerge at the surface. This last
process also includes the tilting of the flux tubes with respect to
the equator. For more information about the models, see the review
by \cite{Charbonneau10}.

Some of these processes are directly observable, in particular, the
surface evolution of the emerged magnetic field, part ({\it{i}}).
Empirically the large-scale evolution is described well by the
surface flux transport (SFT) model \citep[e.g.,][]{Wang89b,
Schrijver01, Mackay02, Baumann04, Jiang10, Mackay12}.  We previously
(CJSS10) used the SFT model to reconstruct the surface field and
open flux for the period 1913--1986. The observed sunspot
longitudes, latitudes, areas, and cycle-averaged tilt angles were
used to create the source term. The results of that model compare
well with the open flux derived from geomagnetic indices
\citep{Lockwood03} and with the reversal times of the polar fields
\citep{Makarov03}.

The remaining processes ({\it{ii}}), ({\it{iii}}), and ({\it{iv}})
deal with the subsurface dynamics, which are currently poorly
constrained by observations. These processes are not included in the
SFT model but are described by the flux transport dynamo (FTD). The
surface fields resulting from the two models are consistent,
provided the FTD model has sufficient near-surface convective
pumping \citep{Cameron12}. The effect of pumping has been studied by
\cite{Karak12}, who demonstrate that the effect of the convective
pumping reduces the memory of the solar activity to one cycle. Other
effects of the pumping on the BL dynamo have also been studied by,
e.g., \citet{Kapyla06}, \citet{Guerrero08},  \citet{DoCao11}, and
\citet{Kitchatinov12}.

Our approach is to study the Babcock-Leighton dynamo model using as
many observational constraints as possible. We thus take the
historical RGO/SOON sunspot record as the basis for constructing the
source of poloidal flux, we employ boundary conditions where the
magnetic field is vertical at photosphere, and we impose substantial
downwards pumping near the surface \cite[as required by
observations,][]{Cameron12}. The differential rotation
\citep{Schou98}, near-surface meridional flow, and near-surface
turbulent diffusivity \citep[see references in][]{Cameron11} are
observationally constrained and modeled in a similar way.

The diffusivity and pumping throughout the bulk of the convection
zone, as well as the deeper structure of the meridional circulation
remain observationally unconstrained. In this paper we first
construct a reference case with intermediate choices (between the
extremes of what can be considered as reasonable) for both the
pumping and the diffusivity in the bulk of the convection zone.
After presenting the results of the calculations for this reference
case, we then consider the effect of varying these parameters. A
more extensive study of the effect of the diffusivity and the
meridional circulation in transporting the field is given by
\cite{Yeates08} for the case without magnetic pumping. The
meridional circulation deep below the surface is controversial, with
opposing views on such basic questions as whether there are one or
two cells in the radial direction. For this paper we have assumed a
one-cell structure.

Here we concentrate on processes  ({\it{i}}), ({\it{ii}}), and
({\it{iii}}), using the FTD model to study the evolution of the
toroidal flux when the source of poloidal flux, process ({\it{i}}),
is based on the historical sunspot record. In a full BL dynamo model
process ({\it{i}}), the appearance on the surface of tilted sunspot
groups, is intimately connected with the formation and rise of flux
loops from the bottom of the convection zone, i.e., process
({\it{iv}}). Since we are using the observed sunspot record for
process ({\it{i}}), we can omit the formation and rise processes
entirely. Ignoring processes ({\it{iv}}) means that we are solving a
driven system rather than a self consistently driven dynamo.
However, it allows us to address the question of whether the polar
field strength at the minimum produced by a BL dynamo with a
poloidal source term derived by observations of sunspot groups and
cycle-average tilt angles is correlated with the observed strength
of the following cycle. To be consistent with observations
\citep[e.g. see][]{Wang09}, a strong correlation should be present
in the model.

The correlation between the minima of the open flux (which is
closely related to the Sun's axial magnetic dipole moment) and the
strength of the subsequent maxima of solar activity
\citep[see][]{Wang09, Jiang11} implies processes ({\it{ii}}) and
({\it{iii}}) are -- on average over a cycle-- essentially linear.
The argument is that a nonlinearity in going between the polar
fields and the toroidal fields at the base of the convection zone
would manifest itself in the number of sunspot groups that appear on
the surface. Such a nonlinearity would then result in maxima of
activity levels that are not strongly correlated with the polar
fields. The maxima of the toroidal field at the base of the
convection zone, which is determined by processes ({\it{i-iii}}),
should therefore also be strongly correlated with the maxima of the
activity level. This is an observationally inferred result, and it
is therefore a test for the model.

There have been two previous efforts to relate the amount of
toroidal flux produced by an FTD model and the observed sunspot
record. The first effort, started by \cite{Dikpati06}, used the time
evolution of the area coverage of sunspots (a measure of the
amplitude of the solar cycle) as input.  A very idealized model was
used to convert this into the source of poloidal flux. In
particular, their source term for cycle $n$ requires knowing of the
timing of the minimum between cycles $n-1$ and $n$. Their finding
was that the toroidal field at the base of the convection zone
during any cycle $n$ in the model was highly correlated with the
actually observed strength of cycle $n$. An explanation for this
correlation is suggested by \cite{Cameron07}: incorporation of the
timing of the minimum in the source term for the poloidal flux, in
conjunction with the Waldmeier effect \citep{Cameron08} introduces
similar correlations even when the amplitudes are independent random
realizations.

The second effort is that of \cite{Jiang07}. They found in the
high-diffusivity regime that the poloidal flux is transported to the
base of the convection zone at the same time as it is transported to
the poles. The simultaneous transport of the field to the poles and
to the base of the convection zone presented in this model offers an
explanation for the observed correlation between the polar fields
and the toroidal field at the base of the convection zone \citep[see
also][]{Choudhuri07}. \cite{Jiang07} then scaled the poloidal field
at the end of each cycle so that the polar fields of the model had
the same strength as those observed by the Wilcox Solar Observatory.
The winding up of this poloidal field by differential rotation
produces a toroidal field that is correlated with the polar fields.
The high correlation between the toroidal fields of the model and
the observed level of activity then follows from the previously
reported correlation between the observed polar fields and the
amplitude of the next cycle \citep{Schatten78}. The exact nature of
the sources in this model is irrelevant, since the field is
rescaled, and this model says little about the causes of the cyclic
amplitude variations.

These previous attempts were carried out before cycle-to-cycle tilt
angle variations had been reported by \cite{Dasi10} and before it
was recognized that a vertical field boundary condition at the
photosphere and strong pumping are required to make the FTD model
match the observations of the surface evolution of magnetic flux
\citep{Cameron12}. The question thus remains open as to what happens
when we use the FTD model with these new constraints. Here we
address this question using the FTD with pumping \citep{Cameron12}
and sources based on as much information as is available (see
CJSS10) to follow the evolution of the subsurface magnetic field
over cycles 15 to 21.

The paper is organized as follows. The BL dynamo model is described
in Section 2. The results for a reference-case set of parameters are
presented in detail in Section \ref{sec:results}. The influence of
the observationally poorly constrained parameters describing the
magnetic diffusivity and pumping in the bulk of the convection zone
is studied in Section \ref{sec:param}. The conclusions and
discussion of our results are given in Section \ref{sec:cons}.

\section{Flux transport dynamo model}

The FTD model is based on the induction equation for an azimuthally symmetric field:

\begin{equation}
\label{eq:At} \frac{\partial A}{\partial
t}=\eta(\nabla^2-\frac{1}{s^2})A-\frac{1}{s}(\boldsymbol{\upsilon'}
\cdot\nabla)(sA)+S,
\end{equation}

\begin{eqnarray}
\label{eq:Bt} \frac{\partial B}{\partial
t}=& &\eta(\nabla^2-\frac{1}{s^2})B+\frac{1}{r}\frac{d\eta}{dr}\frac{\partial}{\partial
r}(rB) \\ \nonumber
&-&\frac{1}{r}\left[\frac{\partial}{\partial
r}(r\upsilon'_rB)+\frac{\partial}{\partial\theta}(\upsilon'_\theta
B)\right]+s(\boldsymbol{B}_p\cdot\nabla)\Omega,
\end{eqnarray}
where
\begin{equation}
\boldsymbol{B}=B(r,\theta)\boldsymbol{e_\phi}
  +\nabla\times[A(r,\theta)\boldsymbol{e_\phi}].
\end{equation}
Here $B(r,\theta)$ is the toroidal component of the magnetic field
and $A(r,\theta)$ the toroidal component of the vector potential,
related to the poloidal component of the magnetic field by
$\boldsymbol{B}_p=\nabla\times(A\boldsymbol{e_\phi})$;
 ${\boldsymbol{\upsilon'}}$ is the sum of the
meridional velocity field ${\boldsymbol{\upsilon}}(r,\theta)$ and magnetic
pumping $\boldsymbol{\gamma}=\gamma(r) {\boldsymbol{e_{r}}}$;
$\Omega(r,\theta)$ is the differential rotation profile; $S(r,\theta,t)$ is the source
term for the poloidal field; $\eta(r)$ is the magnetic
diffusivity; and $s=r\sin\theta$.

The code used to solve the above problem was developed at the MPS.
It has been checked against the benchmark dynamo of \cite{Jouvre08} and was
previously used in \cite{Cameron12}. In essence it treats the
advective term explicitly and an alternating direction implicit scheme
for the diffusive terms. We use 181$\times$71 grid
cells in the latitudinal and radial directions, respectively, and
a timestep of one day.

\subsection{Boundary conditions}
We carry out our calculations in a spherical shell
($0.65R_\odot \leq r \leq R_\odot, 0 \leq \theta \leq \pi$). At the
poles ($\theta=0,~\pi$), we have
\begin{equation}
A=0,~B=0.
\end{equation}
The inner boundary matches a perfect conductor,
\begin{equation}
A=0,~~~~~ \frac{\partial (rB)}{\partial r}=0~~~~~~~\mathrm{at}~r=0.65R_\odot.
\end{equation}
As described in \cite{Cameron12}, the appropriate outer boundary is
that the field is vertical there,
\begin{equation}
\frac{\partial}{\partial r}(rA)=0,~B=0~~~~~~~\mathrm{at}~r=R_\odot.
\end{equation}
The vertical outer boundary condition is also proposed by
\cite{vanBallegooijen07} and has previously been used in the BL
dynamo models by, for example, \cite{MuozJaramillo09} and
\cite{Nandy11}.

\subsection{Flow field and turbulent diffusivity}
We use the same profile for the differential rotation as was used in
\cite{Cameron12} and the references therein. We also use the same
form for the meridional flow as given in that study, except that we
have reduced the parameter $v_0$ that determines the speed of the
meridional flow so that the maximum speed at the surface is now
11~m/s instead of 15~m/s to be consistent with CJSS10.

We also follow \cite{Cameron12} for the turbulent diffusivity profile
\begin{eqnarray}
\label{eq:eta}
\eta(r)=\eta_{rz}
&+&\frac{\eta_{cz}-\eta_{rz}}{2}\left[1+\mathrm{erf}\left(\frac{r-r_{rz}}{d}\right)\right] \nonumber \\
&+&\frac{\eta_{s}-\eta_{cz}}{2}\left[1+\mathrm{erf}\left(\frac{r-r_{s}}{d}\right)\right],
\end{eqnarray}
where the subscript $rz$ is for the radiative zone, $cz$ is for the
convection zone, and $s$ is for the surface properties,
$\eta_{rz}=0.1$~km$^2$ s$^{-1}$, $\eta_{cz}=10$~km$^2$ s$^{-1}$,
$\eta_{s}=250$~km$^2$s$^{-1}$, $r_{cz}=0.7R_\odot$,
$r_s=0.95R_\odot$, and $d=0.02R_\odot$. The argument for the choice
of the surface diffusivity $\eta_{s}$ is given in CJSS10 -- it is
near the middle of the range given by observations \citep[see Table
6.2 of][] {Schrijver00}. This value for $\eta_{s}$ is also
consistent with the range found using comprehensive photospheric
simulations \citep{Cameron11b}. The turbulent diffusivity in the
convective zone $\eta_{cz}$ is not directly constrained. The effect
of different values for this parameter will be studied in
Section~\ref{sec:param}.

In \cite{Cameron12}, constraints were placed on the amount of
magnetic pumping near the top of the convection zone. However, the
strength of the pumping in the deeper layers is unconstrained. We
assume here that the magnetic pumping extends to the base of the
convection zone, albeit at a weaker level than in the near-surface
layer. The radial dependence of the pumping is assumed to be
\begin{eqnarray}
\gamma(r)=
& &\frac{\gamma_{cz}}{2}\left[1+\mathrm{erf}\left(\frac{r-r_{\gamma}}{d}\right)\right] \nonumber \\
&+&\frac{\gamma_{s}-\gamma_{cz}}{2}\left[1+\mathrm{erf}\left(\frac{r-r_{ns}}{d}\right)\right],
\end{eqnarray}
where $\gamma_{cz}=-2$~m~s$^{-1}$, $\gamma_{s}=-20$~m~s$^{-1}$,
$r_{ns}=0.9 R_{\odot}$, and $r_{\gamma}=0.7 R_{\odot}$. We assume
here that the enhanced near surface pumping extends slightly deeper
than the region of enhanced surface turbulent diffusivity
($r_{ns}=0.9 R_{\odot}$ rather than $r_s=0.95 R_\odot$), below which
its strength drops to 2~m/s. Since the value of the pumping in the
convection zone and the depth at which pumping ceases ($r_\gamma$)
are unknown, we study the impact of varying these parameters in
Section~\ref{sec:param}.


\begin{figure*}
\begin{center}
\includegraphics[scale=0.95]{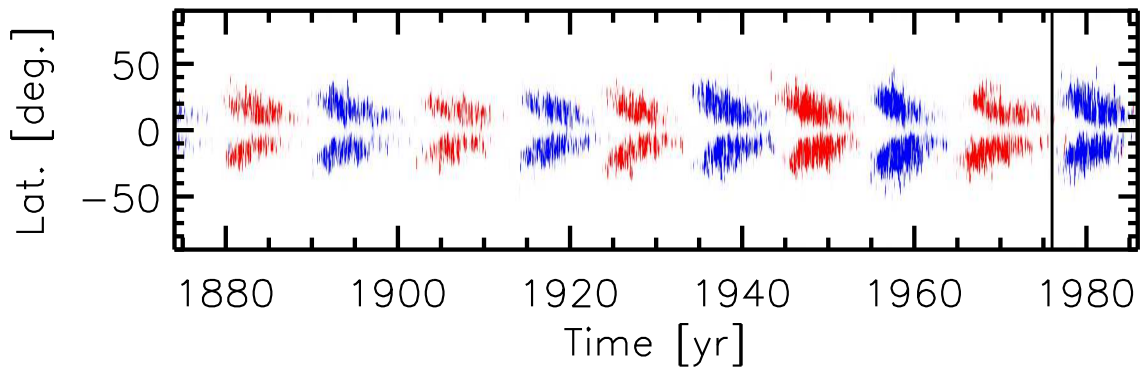}
\end{center}
\caption{Source term, $S(r,\theta,t) \sin(\theta)$ at $r=R_{\odot}$
for the poloidal flux based on the RGO sunspot record. The factor of
$\sin{\theta}$ is included to make the figure proportional to the
amount of signed flux. The red and blue colors indicate positive and
negative signs of $A$, respectively. Data after 1977 is from the
SOON network. } \label{fig:src}
\end{figure*}

\subsection{Source term}
\label{sec:RS_SFT} The source of poloidal field in this model is
based on the observed emergence of bipolar sunspot groups. We use
the procedure of CJSS10 to convert the RGO sunspot group area
observations into a change in the radial magnetic flux at the
surface: the RGO sunspot area record is used to obtain sunspot group
areas, longitudes, and latitudes. Tilt angles from the Mount Wilson
and Kodaikanal observatories \citep{Howard84,Howard91,Sivaraman93}
are averaged over each cycle as in \cite{Dasi10} and CJSS10, and
then used in conjunction with the RGO data for individual spot
groups to determine the longitudinal and latitudinal separation
between the two polarities in the bipolar group. The areas are
converted into fluxes using the parameters in CJSS10 (also see the
references therein). For each sunspot group we calculate the
corresponding change in the radial field at the surface, $\Delta
B_r(R_\odot, \theta, \phi,t)$.

The RGO data covers the period from 1874 to 1976. The tilt angles
are known from 1906 to 1987, and the overlapped data (cycles 14 to
20) can be used to reconstruct the poloidal source term for cycles
15 to 21. The RGO sunspot records were replaced by results from the
SOON network after this date, which used a different definition of
what to include, and therefore introduces a number of
cross-calibration issues. As pointed out in \cite{Cameron12b} there
are outstanding problems when the sunspot record is used in a
non-linear way such as here (the sunspot areas affect the tilt,
which is then multiplied by the area to get the poloidal source
term). We thus restrict our calculation to the period from 1874 to
1981 (i.e., for the cycles for which the toroidal flux at the base
of the convection zone is created from poloidal flux based on the
RGO area data).

The tilt angle data is only available from 1906 onwards, so it can
be used for the sources leading to cycles 15 to 21. For cycles 12
and 13, we use the linear relationship between the sunspot maxima
and the cycle-averaged tilt angles reported by \cite{Dasi10}. While
we restrict our statistical analysis to cycles 15 to 21, simulating
these early cycles makes us less sensitive to the initial condition.

To calculate the change in the azimuthal component of the vector
potential $A$, we then use
\begin{eqnarray}
S(r,\theta,t) \sin{\theta}&=&\frac{R_\odot f(r)}{2 \pi}  \nonumber\\
 &\times & \int_0^\theta \int_0^{2\pi} \Delta B_r(R_{\odot},\theta',\phi,t)
            \mathrm{d} \phi   \sin\theta'\, \mathrm{d}\theta' \nonumber \\
 &\times& \delta(t-t_{\mathrm{em}})
\end{eqnarray}
where $t_{\mathrm{em}}$ is the time at which the sunspot group
was observed to emerge, and $\delta$ is the Dirac delta function.
For the radial dependence of the source term we take
\begin{equation}
f(r)=\frac{1}{2}\left[1+\mathrm{erf}\left(\frac{r-0.9R_\odot)}{d}
\right)\right].
\end{equation}
The resulting sources at the surface are shown in Figure
\ref{fig:src}. This procedure is conceptually related to the
double-ring algorithm developed by \cite{Nandy10}.

\subsection{Initial conditions}
Since we have neither direct measurements of the magnetic fields at
the poles of the Sun in 1874 nor measurements of the toroidal field
of the Sun at the base of the convection zone at any time, the
appropriate initial condition for our model is not well known. The
starting date for our simulations, May 1874, is however about three
fifths of the way through cycle 11 (which ran from March 1867 to
December 1878). This is close to the phase of the cycle where the
axial-dipole field is weak, and given that the minimum of the open
flux was low during this period \citep{Wang09}, $A=0$ is a
reasonable choice. The poloidal component of the initial condition
is particularly important because the decay of the poloidal field is
dominated by the rate at which it is removed from the surface. The
process by which the removal takes place is diffusion across the
equator, which is opposed at the surface by the poleward meridional
flow. The resulting decay rate is about 4000 years (see CJSS10) in
the absence of new sources, so the poloidal component of the initial
condition affects the entire period studied in this paper.

The choice for the initial toroidal field is less critical because
its diffusion rate is determined by diffusion across the equator at
the base of the convection zone. Here the meridional flow is towards
the equator, so it pushes the flux into a thin boundary layer where
the field can easily diffuse across the equator. The pumping plays
an important role here in that it keeps the meridional circulation
from moving the flux upwards and back into the bulk of the
convection zone. For this study we assumed $B=0$ at the start of the
simulation. This may not be adequate, but in view of the short decay
time of the toroidal field it is sufficient to demonstrate that the
Babcock-Leighton model has polar fields at the preceding minima and
toroidal fields at maxima that are strongly correlated with the
maxima of solar activity.

In summary, both the initial toroidal and the initial poloidal field
are zero in our simulations. We use a poloidal source term
constructed from the observed sunspot record rather than the
self-consistent dynamo problem.

\section{Results}
\label{sec:results}

\begin{figure*}
\begin{center}
\includegraphics[scale=.20]{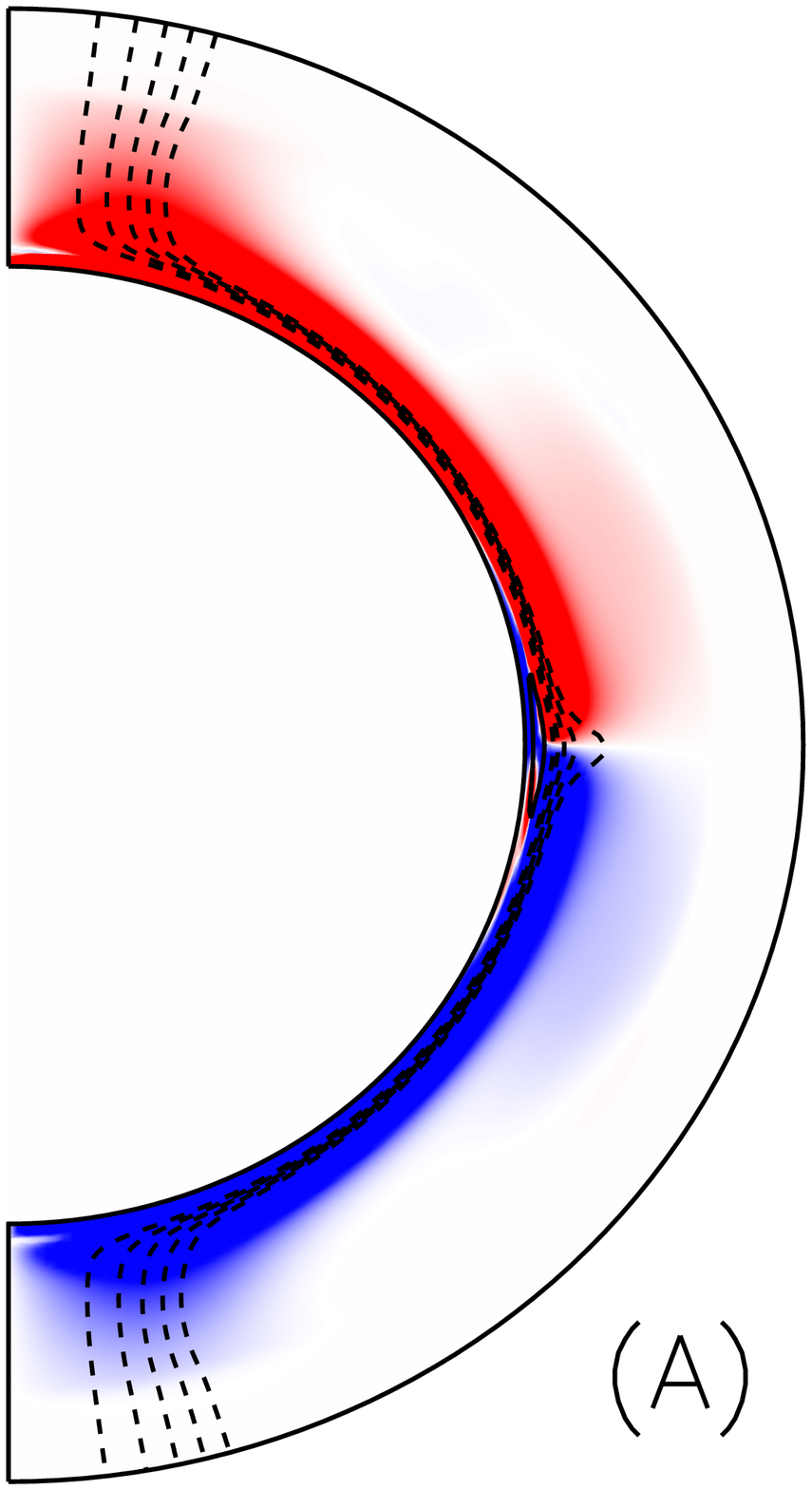}
\includegraphics[scale=.20]{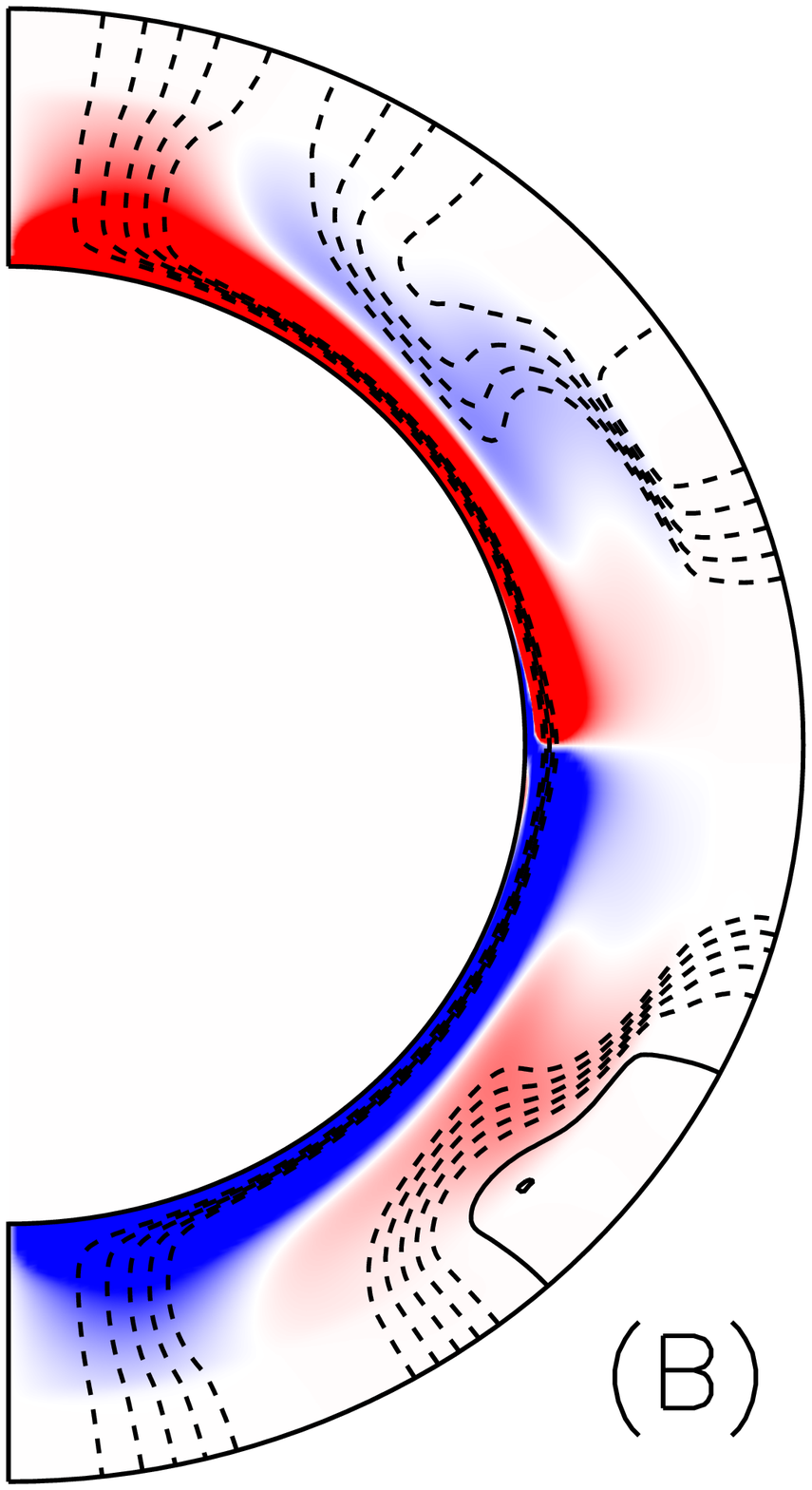}
\includegraphics[scale=.20]{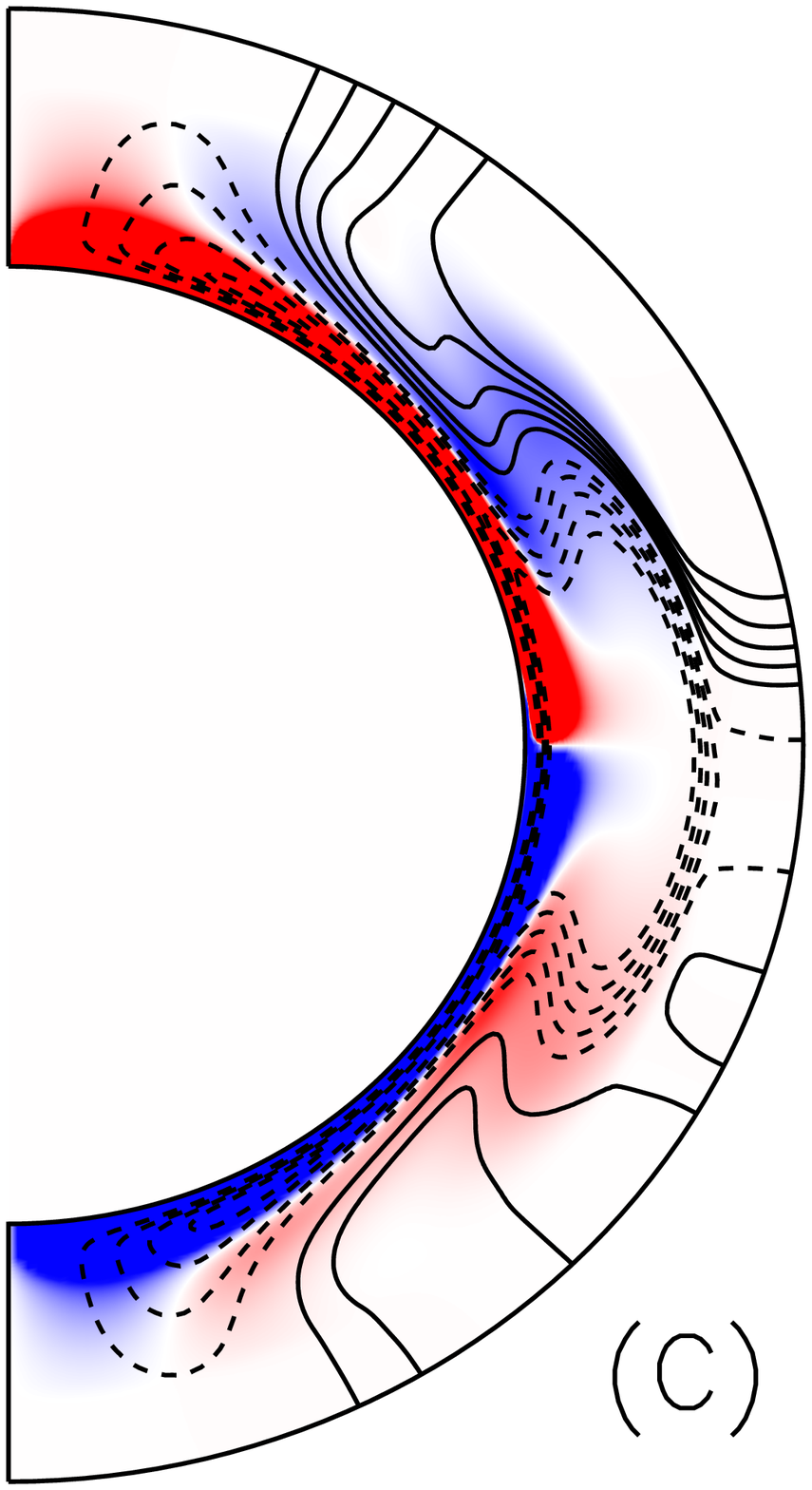}
\includegraphics[scale=.20]{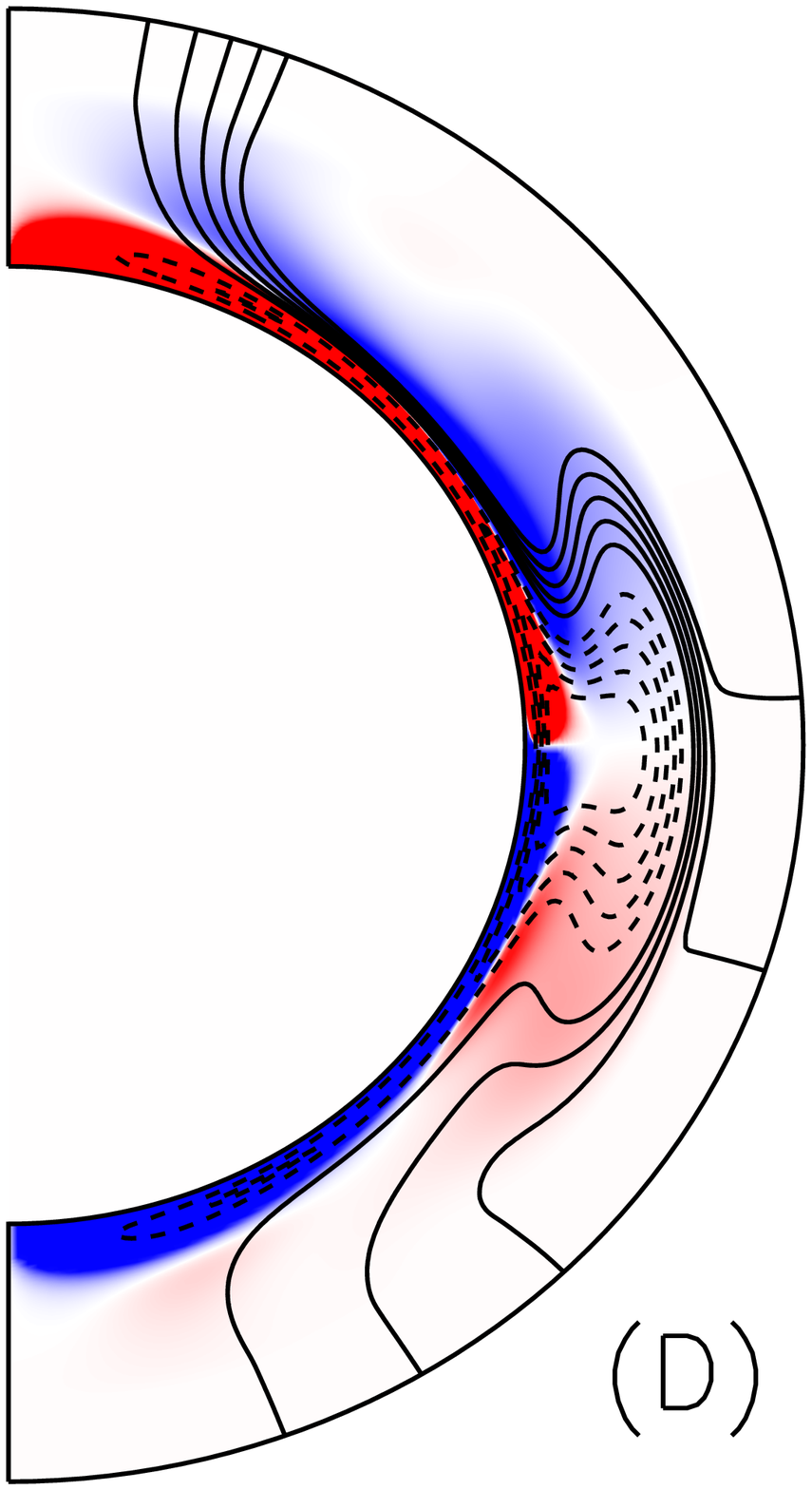}
\includegraphics[scale=.20]{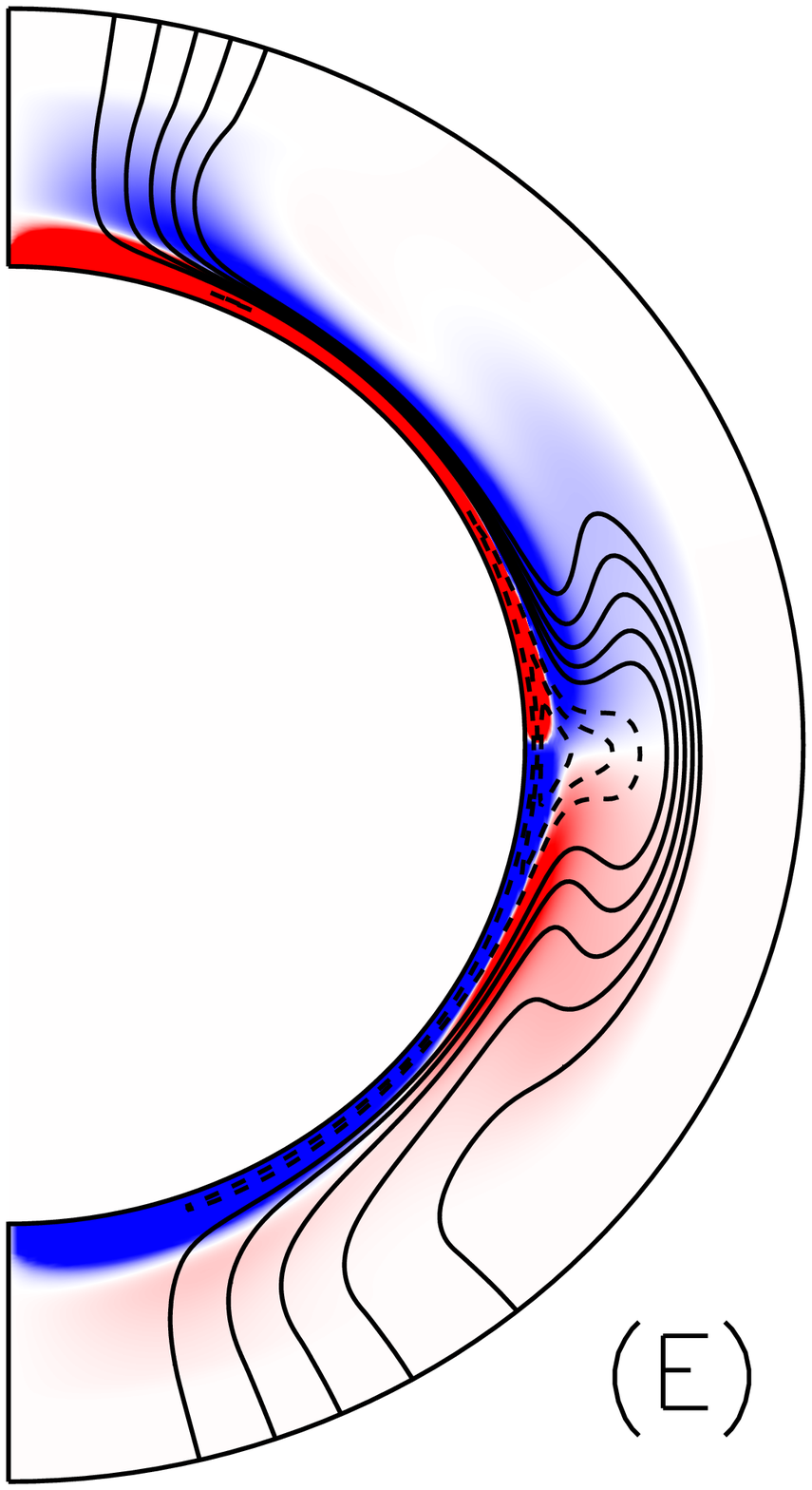}
\includegraphics[scale=.20]{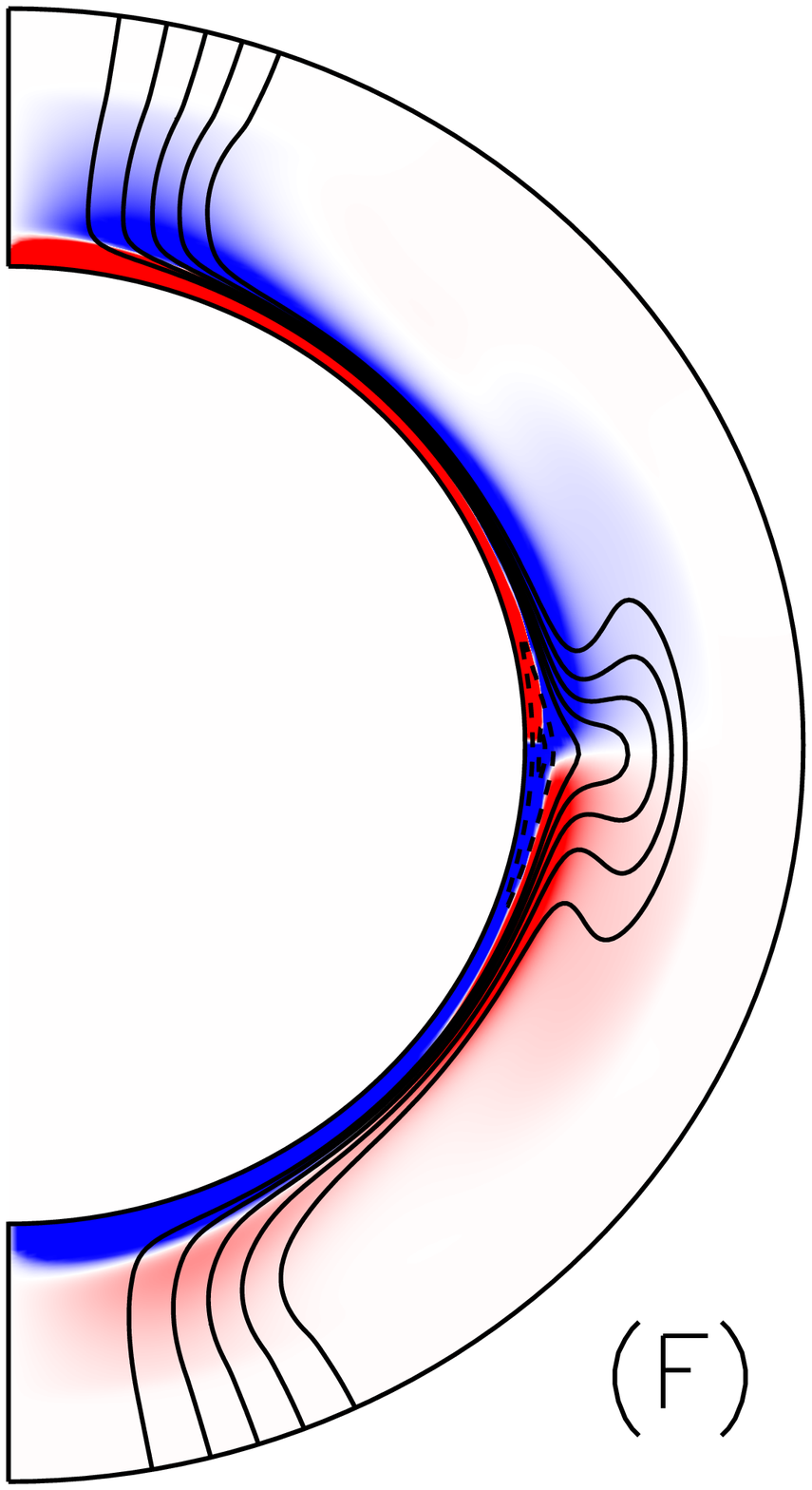}
\caption{Simulated variation of the magnetic field over cycle 19.
The red and blue colors correspond to negative and positive toroidal
fields, respectively. The solid (dashed) lines represent
anti-clockwise (clockwise) orientated field lines of the poloidal
field. The snapshots (A)-(F) cover cycle 19, corresponding to
$t=1956$, $1958$, $1960$, $1962$, $1964$, and $1966$, respectively.}
\label{fig:cyc19}
\end{center}
\end{figure*}

The time evolution of the simulated magnetic field through cycle 19
is shown in Figure~\ref{fig:cyc19} as an example. We see that
throughout the cycle both the poloidal and toroidal fluxes have a
simple form with no `conveyor belt' carrying flux from multiple
cycles \citep[c.f.][]{Dikpati06} operating within the convection
zone, which is consistent with \cite{Karak12}. Although we are in
the `low-medium diffusivity' regime \citep[c.f.][]{Yeates08}, the
transport by convective pumping of poloidal flux from the surface to
the tachocline occurs over the same time span as the flux is
transported to the poles.  The winding up of the field to produce
toroidal flux due to the differential rotation takes place as the
field is descending and after it reaches the base of the convection
zone. The transport by downward pumping near the equator is opposed
by the upward radial flow of the meridional circulation, together
with the equatorward meridional flow, leading to a slower downward
transport near the equator and a delayed reversal of the toroidal
flux near the equator with respect to mid-latitudes.

\begin{figure*}
\begin{center}
\includegraphics[scale=0.95]{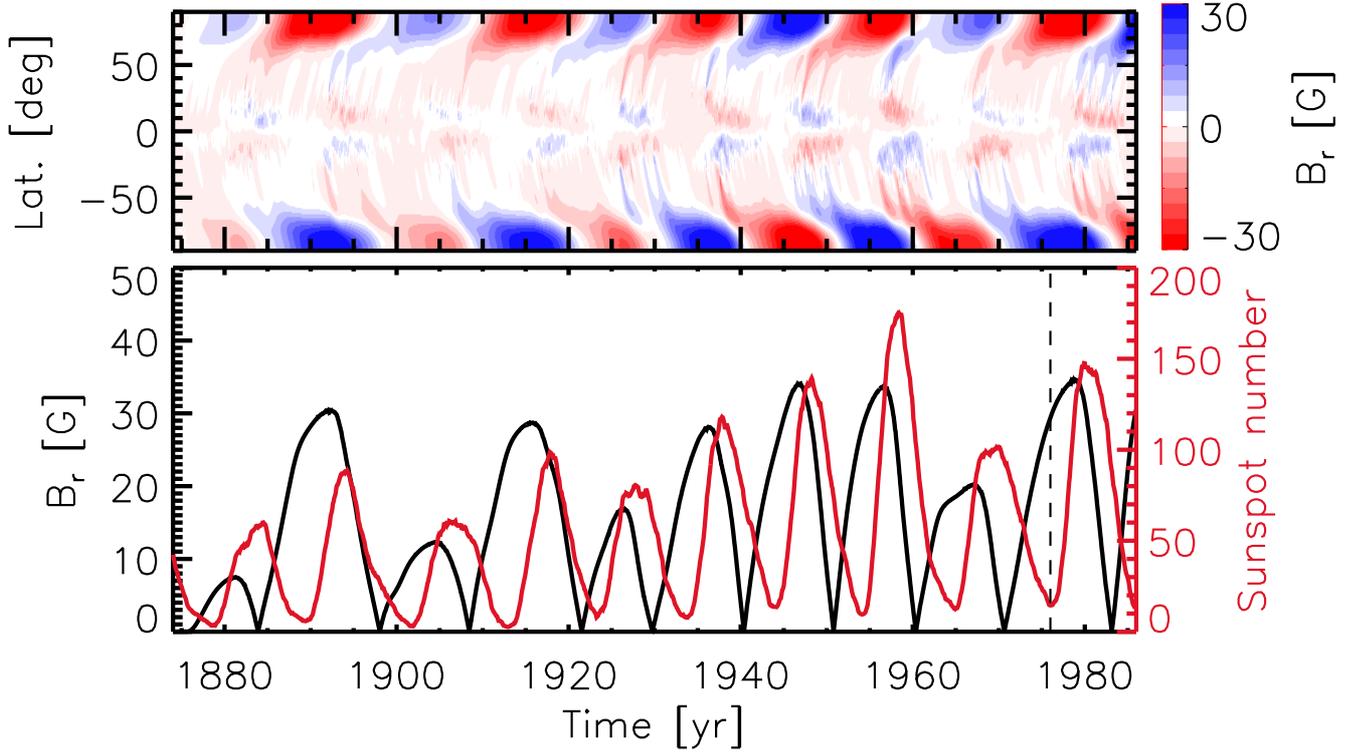}
\end{center}
\caption{Radial component of the surface magnetic field for the
reference case. The upper panel shows a time-latitude diagram. The
lower panel gives the polar fields, defined as the average radial
field strength beyond $\pm 70^{\circ}$ latitude (black curve) and
the sunspot numbers (red curve). The vertical line indicates the
time until which RGO sunspot group area data are available. The
correlation coefficient of the maxima of the polar fields and the
subsequent maxima of sunspot number for cycles 15 to 21 is $0.85$.}
\label{fig:ref_br}
\end{figure*}

The radial component of the field on the surface as a function of
time for the entire period analyzed is shown in
Figure~\ref{fig:ref_br}. The surface evolution of the field is
similar to the one given by the surface flux transport model. Since
with pumping the SFT and FTD models are consistent, and as we are
using the same sources as in CJSS10, the correlation between the
polar fields (here the average field from $\pm70^{\circ}$ latitude
poleward) and the strength of the next cycle (the maxima of the
yearly averaged sunspot number), based on cycles 15 to 21, is
similar to the result in CJSS10 (correlation coefficient $r=0.85$).

\begin{figure*}
\begin{center}
\includegraphics[scale=0.95]{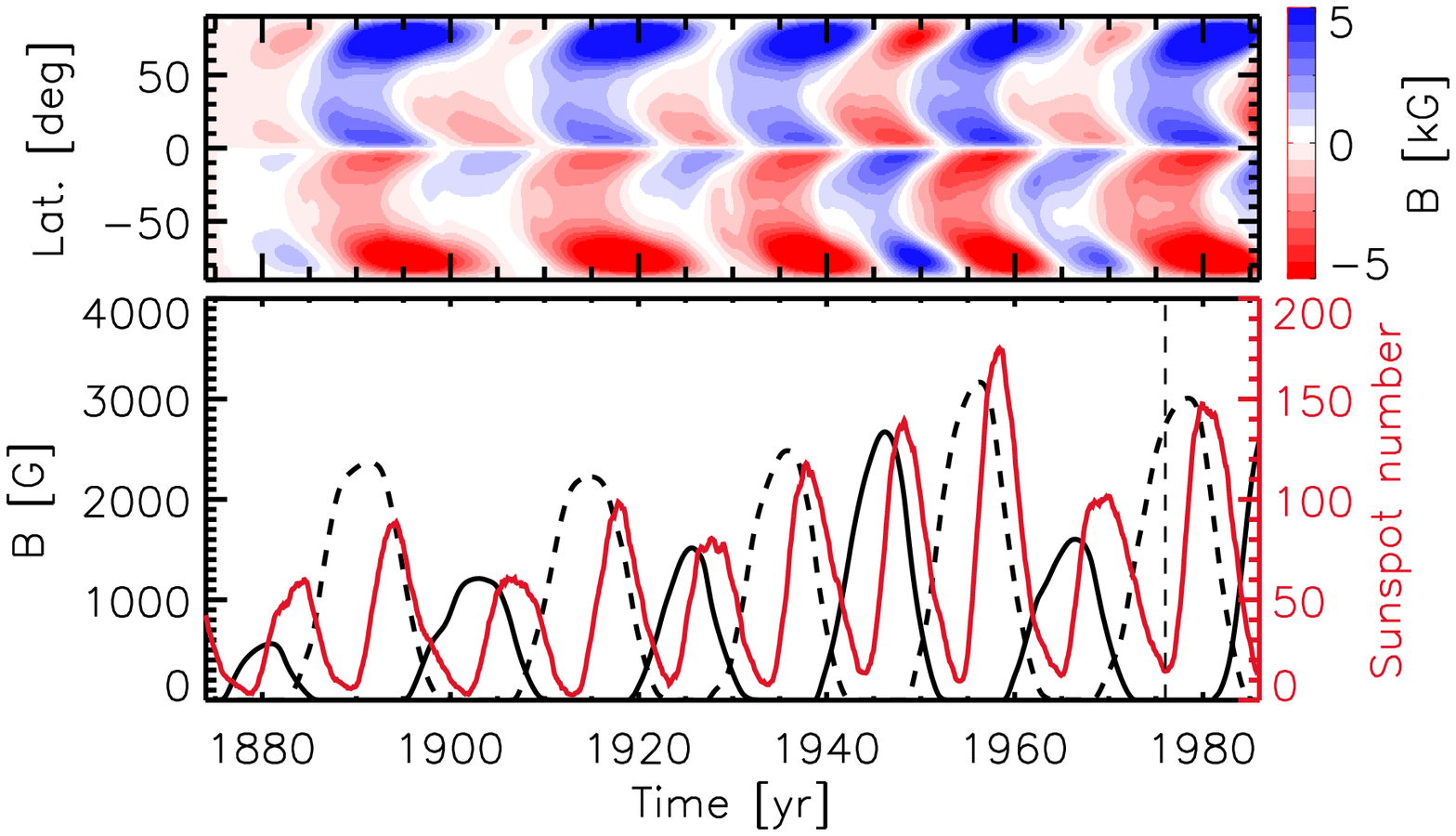}
\end{center}
\caption{Toroidal field at the base of the convection zone
($r=0.7\rsun$). The upper panel shows the evolution as a function of
latitude and time. In the lower panel the average unsigned toroidal
flux between $\pm 45^{\circ}$ latitude corresponding to odd and even
cycles is shown using dashed and solid curves, respectively. The
correlation coefficient between the maxima of the toroidal field at
the base of the convection zone and the maxima of the observed
sunspot number for cycles 15-21 is 0.93.} \label{fig:ref_tor}
\end{figure*}

With our 2-D dynamo model, we can now also consider the evolution of
the toroidal field at the base of the convection zone, which could
not be studied by the SFT model alone. Figure~\ref{fig:ref_tor}
shows the evolution of the toroidal flux at the base of the
convection zone. The toroidal component of the magnetic field at
$r=0.7 R_{\sun}$ is shown in the upper panel. In each hemisphere we
see toroidal flux propagating equatorward at low latitudes. The
reason for this propagation is, in part, the delayed reversal of the
toroidal fields there as discussed above. The eruption of this flux
will result in sunspot groups with an emergence latitude that
migrates equatorward -- that is, with properties similar to the
observed butterfly diagram. There is also a high-latitude branch of
toroidal magnetic field. We conjecture that this branch does not
lead to sunspot emergence because, at latitudes above 75$^{\circ}$,
flux tubes at the bottom of the overshoot layer with field strengths
of $6 \times 10^4$~G or more are stable with regard to the Parker
instability \citep{FerrizMas95, Caligari95}. This flux thus remains
hidden at the base of the convection zone. This explanation
contrasts with those of \cite{Nandy02} and \cite{Guerrero04}, who
show how a meridional flow that penetrates below the tachocline
substantially weakens the high-latitude branch and pushes the
toroidal flux below the tachocline. Flux emergence then can take
place at low latitudes where the meridional circulation causes the
flux to re-enter the convection zone.


The evolution of the unsigned toroidal field is shown in the lower
panel of Figure 4, taking proper account of the overlapping cycles
at the base of the convection zone, at $r=0.7 R_{\sun}$ and averaged
over latitudes $-45^{\circ} < \lambda <45^{\circ}$. The maxima of
the modeled activity levels are clearly related to the observed
amplitude of the cycle, with the correlation coefficient calculated
using cycles 15 to 21 being $r=0.93$.

\section{Effect of the pumping and
               diffusivity parameters in the bulk of the convection zone}
\label{sec:param} 
\begin{figure*}
\includegraphics[scale=0.45]{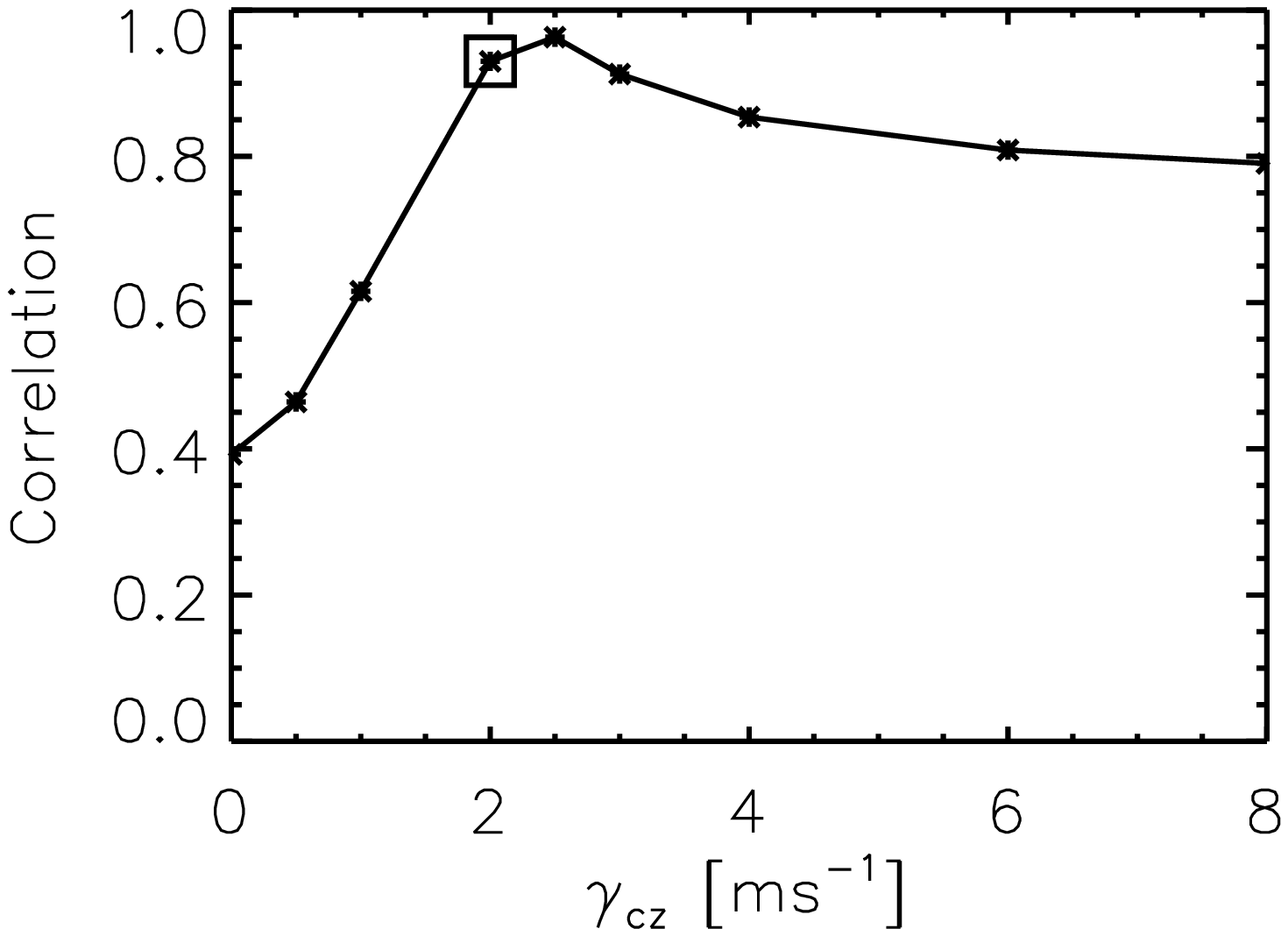}\includegraphics[scale=0.45]{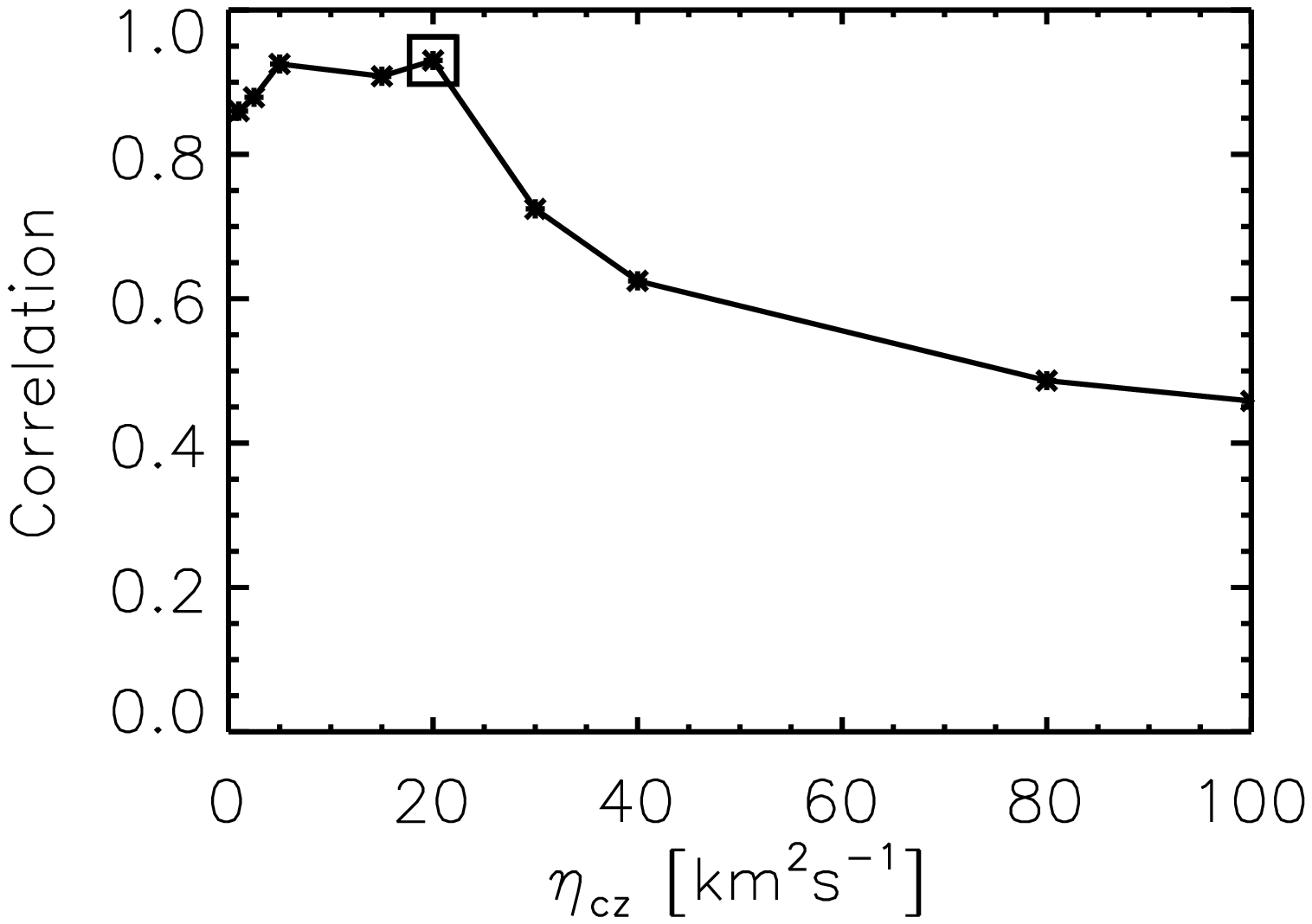}
\caption{Correlation coefficient between the maxima of the unsigned
toroidal field corresponding to each cycle at the base of convection
zone ($r=0.7\rsun$) and the observed maxima of the sunspot number
for simulations with different values of the pumping (left panel),
and the turbulent diffusivity (right panel) in the convection zone.
The asterisk represents the actual numerical experiments which were
performed. The square indicates the result for the reference case. }
\label{fig:par}
\end{figure*}

The dynamics in the bulk of the convection zone
and near the tachocline are much less well constrained by
observations than are the near-surface dynamics. The strength of the
pumping in the bulk of the convection zone ($\gamma_{cz}$) and the
turbulent diffusivity in the convection zone ($\eta_{cz}$) are
therefore free parameters of the model. The evolution of the surface
field is largely independent of these choices, but the flux at the
base of the convection zone is expected to be affected. To
investigate this we performed simulations for which we varied these
two parameters, one at a time. The effect on the correlation between
the toroidal field at the base of the convection zone and the level
of activity is shown in Figure~\ref{fig:par}. The results show that
the good correlation between the toroidal field at the base of the
convection zone and the activity maxima is robust for convective
pumping of 2 to 10 m/s and for diffusivities in the bulk of the
convection zone of 20 km$^2$s$^{-1}$ or less.

\section{Conclusions}
\label{sec:cons}

Our aim was to investigate whether the Babcock-Leighton model, with
sources and parameters based on observations, is consistent with the
correlations implied by the observations. We used a flux transport
dynamo model with pumping and a poloidal source term based on the
historical record of sunspot groups' areas, locations, and tilt
angles to study solar cycles 15 to 21, after starting the
simulations during cycle 11.

Following \cite{Cameron12}, the inclusion of sufficient radial
pumping of the magnetic field means that the surface field from the
flux transport dynamo model, in particular the polar fields, is
similar to the surface flux transport results in CJSS10. In the
present work, we have also been able to study the evolution of the
subsurface field, including the toroidal field at the base of the
convection zone.

We found that the evolution corresponds to a particularly simple
Babcock-Leighton type scenario: in each cycle the poloidal field
from the surface is advected and diffused downwards so that by the
end of a cycle, it completely replaces the previous poloidal flux.
The toroidal field then results from the winding up of this poloidal
field. The memory of the toroidal field is only one cycle
\citep{Karak12}. The decay time of the poloidal field is, in
contrast about 4000 years. The equatorward propagation of the
activity belt is reproduced. We find that the simulated toroidal
flux at the base of the convection zone is highly correlated with
the observed sunspot area observed during the same cycle, consistent
with its being the source of the emerging sunspots. The simulated
minimum of the polar fields between cycles $n-1$ and $n$ are highly
correlated with the maximum activity of cycle $n$. The
Babcock-Leighton model thus appears to be consistent with and able
to explain the observationally inferred correlations.


\begin{acknowledgements}
We are grateful to the referee for very helpful comments on the
paper. We gratefully acknowledge numerous fruitful discussions with
Manfred Sch\"ussler. JJ acknowledges the financial support from the
National Natural Science Foundations of China (11173033, 11178005,
11125314, 11221063, 41174153, 11125314, 2011CB811401) and the
Knowledge Innovation Program of the CAS (KJCX2-EW-T07).
\end{acknowledgements}
\clearpage
\newpage

\bibliographystyle{aa}
\bibliography{BL_Dynamo}

\end{document}